\documentclass[twocolumn,showkeys,preprintnumbers,longbibliography,amsmath,amssymb,epsfig,floatfix,aps,prb]{revtex4-2}
\usepackage[table, svgnames]{xcolor}
\usepackage{xcolor}
\usepackage{tabu}
\usepackage{amsmath}
\usepackage{mathtools}
\usepackage{physics}
\usepackage[version=4]{mhchem}
\usepackage[english]{babel}
\usepackage[utf8]{inputenc}
\DeclareMathAlphabet{\altmathcal}{OMS}{cmsy}{m}{n}
\usepackage{siunitx}
\usepackage{multirow}
\usepackage{amsfonts}
\usepackage{amssymb}
\usepackage{calligra}
\usepackage{calrsfs}
\DeclareMathAlphabet{\mathcalligra}{T1}{calligra}{l}{m}
\usepackage{epsfig}
\usepackage{graphicx}
\usepackage{dcolumn}
\usepackage{color}
\usepackage{natbib}  
\usepackage{hyperref}
\usepackage{breakurl}
\hypersetup{colorlinks=true, citecolor=blue, urlcolor=blue, linkcolor=blue}
\usepackage{bm}
\usepackage{booktabs}
\usepackage{mathtools}

\usepackage{tabularray}
\usepackage{booktabs}
\newcolumntype{L}[1]{>{\raggedright\arraybackslash}p{#1}}
\newcolumntype{P}[1]{>{\centering\arraybackslash}p{#1}}
\newcolumntype{R}[1]{>{\raggedleft\arraybackslash}p{#1}}

\usepackage[table]{xcolor} 
\definecolor{LightCyan}{rgb}{0.88,1,1}
\usepackage{graphicx}      
\usepackage{dcolumn}       
\usepackage{ifthen}

\usepackage{amsmath}
\usepackage{mathtools}
\usepackage{amsfonts}
\usepackage{amssymb}
\usepackage{physics}       
\usepackage{bm}            
\usepackage{calrsfs}
\DeclareMathAlphabet{\altmathcal}{OMS}{cmsy}{m}{n}
\DeclareMathAlphabet{\mathcalligra}{T1}{calligra}{l}{m}

\usepackage[utf8]{inputenc}
\usepackage[english]{babel}
\usepackage{ulem}
\usepackage{siunitx}
\usepackage{multirow}
\usepackage{booktabs}

\begin{document}
\title{Gate-tunable giant anomalous Hall effect in magnetic topological insulator bilayer}
\author{Basavaraja G} 
\author{Mukul Kabir}
\email{mukul.kabir@iiserpune.ac.in}
\affiliation{Department of Physics, Indian Institute of Science Education and
Research, Pune 411008, India}

\begin{abstract} 
In the two-dimensional limit, the intrinsic magnetic topological insulator \ce{MnBi2Te4} provides a compelling platform for exploring thickness-dependent quantum states and their evolution under external perturbations.  %
Using first-principles calculations and classical Heisenberg Monte Carlo simulations, we demonstrate that electrostatic gating and surface chemical functionalization can drive a systematic crossover from the topological to the conventional anomalous Hall regime. This transition is governed by the simultaneous shift of the Fermi level away from the topological gap and a reversal of interlayer coupling from antiferromagnetic to ferromagnetic order. Results reveal that hole doping drives the Fermi level into the valence bands, inducing an exceptionally high anomalous Hall conductivity of 1127 S/cm arising from Berry curvature hot spots. In contrast, surface chemical doping drives a topological state where intrinsic $\sigma_{xy}$ is reduced from $e^2/h$ to $\sim 0.86\ e^2/h$ by the spectral conexistance of chiral edge mode with metallic bulk states of the two-dimensional film. Furthermore, we show that both tuning routes significantly enhance in-plane exchange interactions, leading to a substantial increase in the magnetic ordering temperature relative to the pristine bilayer. %
These results establish a versatile framework for the simultaneous engineering of topological, magnetic, and transport properties in ultrathin \ce{MnBi2Te4}, offering direct implications for the development of reconfigurable quantum devices.
\end{abstract}
\maketitle

\section{Introduction} 
Topological quantum materials have emerged as a significant area of research in condensed matter physics due to their unique electronic properties and potential applications in advanced technologies. These materials, including topological insulators, Dirac and Weyl semimetals, and topological superconductors, exhibit robust surface states that are protected by topological invariants~\citep{nphys4302, nphys4274, nmat5012, yan2017topological, RevModPhys.90.015001}. This topological protection endows them with remarkable resilience to defects and impurities, making them promising candidates for quantum computing and spintronics applications. 

While the bulk of topological insulators is gapped, they host dissipationless edge currents that are immune to backscattering, a phenomenon arising from the interplay of spin-orbit (SO) coupling and time-reversal (TR) symmetry~\citep{PhysRevLett.95.146802,science.1148047,PhysRevLett.98.106803,nature08916,hasan2010colloquium}. These surface states harbour massless Dirac fermions exhibiting spin-momentum locking, where the electrons with opposite spins propagate in opposite directions. The introduction of spontaneous magnetization breaks TR symmetry, causing the Dirac fermion to acquire mass and open an exchange gap in the Dirac surface band.  Tuning the Fermi level within this exchange gap results in the emergence of the quantum anomalous Hall (QAH) effect, which manifests without an external magnetic field~\citep{tokura2019magnetic}.

Engineered topological insulators have emerged as promising materials in the exploration of topological quantum phenomena~\citep{tokura2019magnetic,chang2016quantum,he2018topological}. 
In topological insulators doped with magnetic ions such as \ce{Cr}- and \ce{V}-doped \ce{(Bi{,}Sb)2Te3}, the QAH effect is observed when doping induces long-range magnetic order and aligns the Fermi level within the exchange gap~\citep{science.1234414,chang2015high}. However, the QAH effect in these systems is limited to extremely low temperatures, typically in the millikelvin range, due to material imperfections and magnetic disorder. In contrast, intrinsic magnetic topological insulators exhibit a more robust QAH effect at higher temperatures, potentially operable above the temperature of liquid helium~\citep{otrokov2017highly, otrokov2019prediction, gong2019experimental, he2020mnbi2te4}. The intricate interplay between symmetry, magnetism, and topology in these materials gives rise to a diverse array of topological quantum states. By adjusting parameters such as external magnetic fields, temperature, and dimensional confinement, these materials can transition between different phases, including QAH insulators, axion insulators, quantum spin Hall insulators, and Dirac or magnetic Weyl semimetals~\citep{PhysRevLett.120.056801, deng2020quantum, PhysRevLett.122.206401,liu2021magnetic,liu2020robust,  wang2021intrinsic, he2020mnbi2te4,li2019intrinsic,10.1063/5.0015328, ge2020high}.

\ce{MnBi2Te4} serves as a quintessential example of an intrinsic magnetic topological insulator~\citep{otrokov2017highly, otrokov2019prediction, gong2019experimental,PhysRevLett.122.206401}, exhibiting A-type antiferromagnetism with a N\'eel temperature of 25 K, where ferromagnetic layers stack antiferromagnetically~\citep{PhysRevLett.122.107202,PhysRevLett.124.167204,PhysRevB.99.155125,PhysRevResearch.1.012011,PhysRevX.9.041038,PhysRevX.11.011003}. The magnetism can be controlled to manifest various topological quantum phases by tuning external magnetic and electric fields~\citep{deng2020quantum,liu2020robust,PhysRevB.103.104403}, applying strain~\citep{PhysRevB.101.184426,acs.nanolett.1c01874,acs.nanolett.5c05229}, inter-layer sliding~\citep{PhysRevB.106.235302}, twisting~\citep{PhysRevLett.124.126402}, alloying~\cite{PhysRevB.103.064412}, intercalation~\cite{article_127842}, and adjusting thickness~\citep{deng2020quantum,liu2020robust,PhysRevLett.122.107202}. Remarkably, the van der Waals (vdW) nature of \ce{MnBi2Te4} allows exfoliation into two-dimensional (2D) layers, offering significant tunability. In thin films with an odd number of layers, the perpendicular antiferromagnetic order does not completely cancel the magnetic moments, enabling the QAH effect under appropriate conditions~\citep{deng2020quantum}. In contrast, even-layer \ce{MnBi2Te4} exhibits no net magnetic moment and forms an axion insulator characterized by a gapped topological surface state~\citep{PhysRevLett.122.206401,liu2021magnetic}. Furthermore, a quantum phase transition between QAH and axion insulators has also been observed at high magnetic fields, exceeding 9 T~\citep{liu2020robust}.

The responses of these topologically protected quantum states to external perturbations remain a central yet largely unexplored question. In \ce{MnBi2Te4}, a particularly revealing perturbation is shifting the Fermi level $E_{\rm F}$ away from the exchange gap into the bulk bands. This crossover to a metallic regime fundamentally alters the electronic landscape. As bulk states become available at $E_{\rm F}$, they provide channels for backscattering, and the topological protection of the insulating states is lost. In this regime, the Berry curvature integral is no longer evaluated over the full Brillouin zone but is instead bounded by the Fermi surface. Consequently, the anomalous Hall conductivity is no longer pinned to the Chern number of fully occupied bands, and a non-quantized response emerges that depends sensitively on the Fermi surface. Continuous tuning of  $E_{\rm F}$ across this crossover, accessible via electrostatic gating or controlled defect engineering, thus provides a unique pathway to track the evolution from the topological to the conventional anomalous Hall regime, with direct implications for reconfigurable Hall devices. Despite its significance, a systematic investigation of this crossover in \ce{MnBi2Te4} remains elusive.  Beyond Fermi level tuning, the interlayer antiferromagnetic exchange coupling provides an additional control parameter.  Suppressing or reorienting this coupling modifies the net out-of-plane magnetization and the Berry curvature distribution near the topological gap. As a result, it may potentially produce a large anomalous Hall effect in the 2D limit, even outside the quantized regime.

In this study, we employ strain, electrostatic gating, and surface chemical doping as controlled perturbations to manipulate the electronic and magnetic states of an ultrathin \ce{MnBi2Te4} flake of fixed thickness. We focus on bilayer \ce{MnBi2Te4}, which is an axion insulator with compensated interlayer magnetism. Through first-principles calculations, we demonstrate that both the insulating state and the interlayer antiferromagnetic (AFM) coupling can be altered by electrostatic gating and surface chemical doping. Notably, hole doping shifts the Fermi level into a partially filled band and induces an exceptionally high anomalous Hall conductivity (AHC), driven by Berry curvature. This represents a significant enhancement of the AHC reaching as high as 1127 S/cm, exceeding known 2D magnets~\citep{s41563-018-0132-3}, and \ce{MnBi2Te4} thin films~\citep{s41467-022-29259-8}.  
 Surface chemical doping, on the other hand, transforms the system into a metallic ferromagnetic phase, where quantization of the Hall conductivity is partially suppressed due to backscattering from bulk states at $E_{\rm F}$. We further compute the exchange interactions and investigate the resulting magnetic ordering through classical Heisenberg Monte Carlo simulations. The results reveal a more than two-fold enhancement of the magnetic ordering temperature in the doped bilayers. Collectively, these findings establish new pathways for tuning the electronic, magnetic, and topological properties of ultrathin \ce{MnBi2Te4} sheets.

The remainder of the manuscript is organized as follows. Section~\ref{methods} outlines the computational methodology, and its validity is established through comparison with available experimental results in Section~\ref{resultsA}. Section~\ref{resultsB} examines how electrostatic gating and surface chemical doping modify the electronic, magnetic, and topological properties of the bilayer.  Discussion includes the microscopic mechanisms underlying these changes, including the emergence of colossal and incompletely quantized AHC. Finally, Section~\ref{summary} summarizes the main findings and provides an outlook.

\section{\label{methods} Methods}
Electronic structure calculations were performed using density functional theory (DFT) as implemented in the VASP code~\citep{PhysRevB.47.558,PhysRevB.54.11169}. The wavefunctions were represented using the projector-augmented wave method~\citep{PhysRevB.50.17953}, with a kinetic energy cutoff of 500 eV applied for the plane-wave expansion. In the Dudarev scheme~\citep{PhysRevB.57.1505}, an on-site Coulomb interaction of $U_{\rm Mn}=4$ eV is added to the exchange-correlation energy described by the Perdew-Burke-Ernzerhof functional~\citep{PhysRevLett.77.3865}.  The vdW interaction between the septuple-layers (SLs) is described using the DFT-D3 method~\citep{10.1063/1.3382344}. All structures are optimized until all force components are reduced below a threshold of 0.005 eV/\AA. Both structural optimizations and static electronic structure calculations are performed using a $\Gamma$-centred Monkhorst-Pack $k$-grid.~\citep{PhysRevB.13.5188} For the 2D Brillouin zone, a grid of 15$\times$15$\times$1 is used, while for the 3D Brillouin zone, a grid of 15$\times$15$\times$5 is employed for in-plane (1$\times$1) cells for the 2SL nanosheet and bulk, respectively. The 2SL \ce{MnBi2Te4} structures are simulated as periodic films, separated by a vacuum gap of at least 20 \AA. This ensures that interactions between the periodic images are negligible. Spin-orbit coupling is self-consistently incorporated in all calculations. The tight-binding calculations are conducted by interfacing the VASP code~\citep{PhysRevB.47.558,PhysRevB.54.11169}  with the WANNIER90 package~\citep{PhysRevB.56.12847,MOSTOFI2008685} and WannierTools~\citep{wu2018wanniertools}. For the bilayer, the {\em bulk} states denote the electronic band manifold under in-plane periodic boundary conditions, in the context of bulk-boundary correspondence. The boundary states are calculated using the iterative Green's function method, yielding the boundary spectral function of the semi-infinite system, which are surface states for a 3D geometry and edge states for the 2D bilayer film.

The anomalous Hall conductivity is calculated as the Brillouin zone (BZ) integral of the total Berry curvature $\Omega_{\alpha\beta}$,
\begin{equation}
\sigma_{\alpha\beta} = -\frac{e^2}{\hbar}\int_{\rm BZ} \frac{d\mathbf{k}}{(2\pi)^3} \Omega_{\alpha\beta}(\mathbf{k}),
\label{eq1}
\end{equation}
where $\Omega_{\alpha\beta}(\mathbf{k})$ is the total Berry curvature.
Applying the Kubo formula~\citep{PhysRevLett.49.405,PhysRevLett.92.037204}, the total Berry curvature is evaluated using the tight-binding Hamiltonian in the Wannier basis~\citep{PhysRevB.56.12847},  following the
occupation-difference formulation~\citep{PhysRevB.74.195118} as implemented in WannierTools~\citep{wu2018wanniertools},
\begin{eqnarray}
\Omega_{\alpha\beta}(\mathbf{k}) &=& -\Im \sum_{n \neq m} \left[f_n(\mathbf{k}) - f_m(\mathbf{k})\right] \nonumber \\
&\times&\frac{\langle u_{n\mathbf{k}}|\nabla_{\alpha}H(\mathbf{k})|u_{m\mathbf{k}}\rangle \langle u_{m\mathbf{k}}|\nabla_{\beta}H(\mathbf{k})|u_{n\mathbf{k}}\rangle}{[E_n(\mathbf{k}) - E_m(\mathbf{k})]^2}. 
\end{eqnarray}
Here $\nabla_{\alpha}H$ is the velocity operator, $|u_{n\mathbf{k}}\rangle$ and $|u_{m\mathbf{k}}\rangle$ are cell-periodic Bloch states for each $\mathbf{k}$ with energies $E_n(\mathbf{k})$ and $E_m(\mathbf{k})$, respectively. $f_n$ is the Fermi--Dirac occupation function. Because each term carries the occupation difference $(f_n - f_m)$, pairs of states with equal occupation, specifically those made degenerate by $\altmathcal{PT}$ symmetry, which share identical energies and occupations, yield a vanishing contribution. Consequently, the expression remains well-defined and finite at these degeneracy points. The corresponding Chern number is calculated as Berry flux over the 2D Brillouin zone, $\altmathcal{C} = 1/(2\pi)\int_S \Omega_z(\mathbf{k})dk^2$, where $S$ is the area of the 2D BZ and $\Omega_z$ is the normal component of the Berry curvature.

For a system with partially filled bands, $E_{\rm F}$ restricts the integration domain such that $\int_{\rm BZ} \rightarrow \sum_n \int_{E_n(\mathbf{k})\leqslant E_{\rm F}}$ in Eq.~\ref{eq1}. 
Consequently, $\sigma_{xy}$ is evaluated over the Fermi sea rather than the full BZ. In metallic systems, or when the chemical potential is tuned away from the topological gap, $\sigma_{xy}$ depends sensitively on the exact position of $E_{\rm F}$, resulting in a non-quantized response for the partially filled bands. This behavior can be further understood through the Kubo decomposition, 
$\sigma_{xy} = - (e^2/h)\int_{-\infty}^{E_{\rm F}} dE D(E) \bar{\Omega}(E)$, 
where $D(E)$ is the density of states, and $\bar{\Omega}(E)$ is the Berry curvature averaged over the iso-energy surface at energy $E$. The Hall conductivity can therefore be written as, 
$\sigma_{xy} = -(e^2/h) \altmathcal{C} + \delta \sigma_{xy}(E_{\rm F})$, 
where the first term represents the contribution from the bands below the topological gap, and the second term results from the partially filled metallic band, which destroys the transport quantization and continuously varies with $E_{\rm F}$.

To describe the magnetism and explore the magnetic ordering, we employ a long-range anisotropic Heisenberg Hamiltonian defined on the stacked triangular lattice of localized \ce{Mn} moments~\citep{PhysRevB.103.214411,PhysRevMaterials.6.084407}, 
 \begin{eqnarray}
\altmathcal{H} = &-& \frac{1}{2}\left[\sum_{\mathclap{{\langle i, j\rangle}}}J_{1}\bm{S}_i \cdot \bm{S}_j +  \sum_{\mathclap{{\langle \langle i, j\rangle\rangle}}} J_{2}\bm{S}_i \cdot \bm{S}_j + \sum_{\mathclap{{\langle \langle\langle i, j\rangle\rangle\rangle}}} J_{3}\bm{S}_i \cdot \bm{S}_j \right]\nonumber \\ 
&-& \frac{1}{2}\sum_{\mathclap{{\langle i, j\rangle_{\perp}}}} J_{\perp}\bm{S}_i \cdot \bm{S}_j - \sum_i A_z S_i^z S_i^z, 
\end{eqnarray}
where $J_1$, $J_2$, $J_3$ denote isotropic first, second, and third neighbor in-plane exchange interactions, respectively. $J_{\perp}$ is the interlayer exchange interaction. The interaction is ferromagnetic for positive $J$ values and antiferromagnetic for negative $J$ values. $A_z$ represents the on-site single-ion anisotropy, where $A_z > 0$ ($A_z < 0$) denotes easy-axis (easy-plane) magnetism. We considered various spin-ordered phases, including ferromagnetic (FM), zigzag antiferromagnetic (Z-AFM), stripe antiferromagnetic (S-AFM), and double-stripe antiferromagnetic (DS-AFM) configurations to compute the isotropic exchange interactions through energy mapping. Magnetic exchange interactions are calculated using larger in-plane ($\sqrt{3}\times2$) and ($2\sqrt{3}\times2$) supercells. The single-ion anisotropy is calculated based on the energy difference between in-plane $[100]$ and out-of-plane $[001]$ magnetization, $A_z = (E{\mathrm [100]} - E{\mathrm [001]})/S^2$. The corresponding magnetic ordering temperature is evaluated using the classical Heisenberg Monte Carlo method~\citep{Toth_2015}.

\begin{figure*}[!t] 
\centering 
\includegraphics[scale=0.27]{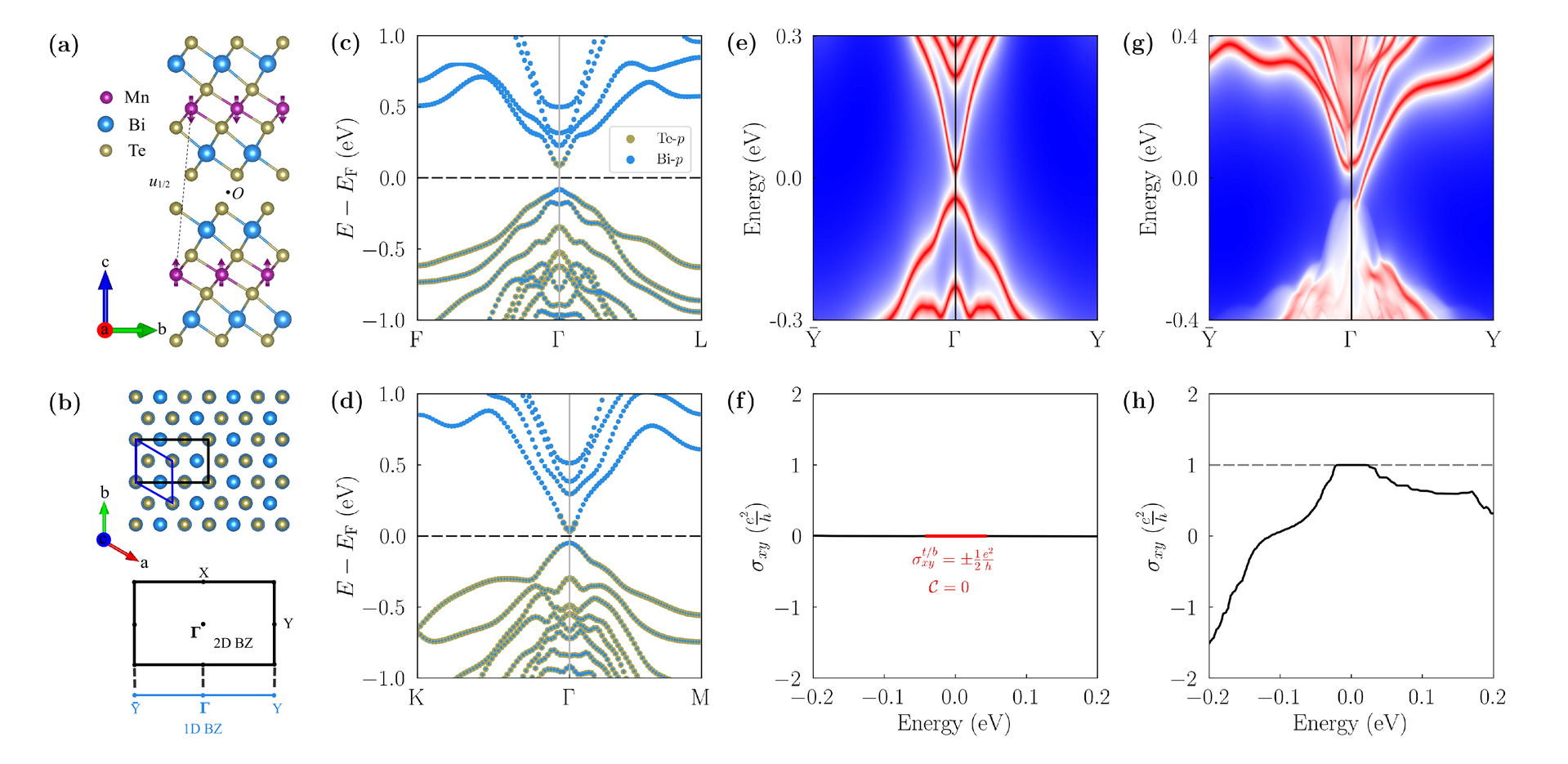}
\caption{
 Crystal structure and topological phases of \ce{MnBi2Te4}. 
(a) The schematic crystal structure of \ce{MnBi2Te4} illustrates ferromagnetic \ce{Mn} layers that are coupled antiferromagnetically across the vdW gap, exhibiting A-type AFM ordering. The vector $\mathbf{u}_{1/2}$ represents the half-translation connecting these \ce{Mn} layers, while $O$ indicates the inversion center of the structure. (b) The top view along the $c$-axis highlights the hexagonal unit cell, outlined by solid blue lines, alongside the solid black rectangular supercell employed for surface state calculations. Additionally, it shows the two-dimensional BZ and the projected one-dimensional BZ, along which the edge modes are computed. The projected band structures for (c) the bulk and (d) the bilayer exhibit band inversion, confirming the existence of a topological insulator state. (e) - (f) The bilayer with antiferromagnetic interlayer coupling presents a gapped boundary (edge) spectrum and a zero-plateau anomalous Hall conductivity ($\sigma_{xy}^{\rm t/b} = \pm \frac{1}{2} e^2/h, \altmathcal{C}=0$), characteristics indicative of an axion insulator. (g) - (h) A topological edge mode emerges for the ferromagnetic bilayer, providing a dissipationless conducting channel that results in a quantized Hall conductivity.
} 
\label{fig:fig1} 
\end{figure*}
   
\section{\label{results}Results and Discussion}
We begin by comparing bilayer \ce{MnBi2Te4} with its bulk counterpart,  setting the stage for the remainder of the manuscript. The good agreement with available experimental data validates the present computational approach. We then proceed to explore how the electronic, magnetic, and topological properties of the bilayer evolve under various external perturbations.  

\subsection{\label{resultsA}Insulating bulk and bilayer \ce{MnBi2Te4}}
The crystal structure of \ce{MnBi2Te4} is closely related to tetradymite topological insulators like \ce{Bi2Te3}, which consist of stacked quintuple \ce{Te-Bi-Te-Bi-Te} layers. \ce{MnBi2Te4} crystallizes in a rhombohedral structure with space group $R\bar{3}m$, consisting of  \ce{Te-Bi-Te-Mn-Te-Bi-Te} septuple layers, held together by vdW interaction [Figure~\ref{fig:fig1}(a) and \ref{fig:fig1}(b)].  The optimized lattice parameters for the bulk, $a=$ 4.37 and $c=$ 40.28 \AA\, agree well with the experimental values~\citep{lee2013crystal,zeugner2019chemical,PhysRevMaterials.3.064202}. In the bilayer, the in-plane lattice parameter contracts slightly to 4.36 \AA.

\begin{table}[b]
\caption{Magnetic exchange interactions are calculated using first-principles methods. In both bulk and bilayer \ce{MnBi2Te4}, the in-plane magnetism is primarily governed by FM first-neighbor interaction, $J_1$, while $J_2$ and $J_3$ play a crucial role in the doped bilayers. The calculated magnetic ordering temperatures, $T_{\rm C/N}$, for both bulk and bilayer systems, are in good agreement with experimental observations~\citep{PhysRevX.11.011003,otrokov2019prediction}. Notably, the interlayer magnetic coupling, $J_\perp$, transitions to ferromagnetic in the 0.2 holes/f.u. hole- and \ce{Br}-doped systems.}
\begin{tblr}{
  colspec={Q[l, 2.5cm] X[0.7, c] X[0.7, c] X[0.7, c] X[0.7, c] X[0.7, c] X[0.7, c]},
  rowsep = 0pt,
}
\hline
\hline
System & \SetCell[c=5]{c} Exchange interactions (\si{\micro\electronvolt}) &&&&& $T_{\rm C/N}$ \\
                                               & $J_1$ & $J_2$  & $J_3$ & $J_{\perp}$ & $A_z$ &  (\si{\kelvin})  \\
\hline      
        Bulk \ce{MnBi2Te4}            & 231 & $-$18 & 4 & $-$40      & 184 & 31.0 \\
        2SL                                  & 252 & $-$13 & 4 & $-$36      & 55 & 25.5 \\
        hole-doped 2SL                 & 453 &  38 &   25 &  153      & 25 & 60.0 \\
        Br-doped 2SL                    & 270 &  126 &  46 &  88      & 28 & 54.0 \\ 
\hline
\label{tab:tab1}
\end{tblr}
\end{table}

The non-trivial electronic topology arises from the inverted \ce{Bi-Te} bands [Figure~\ref{fig:fig1}(c) and \ref{fig:fig1}(d)], while the triangular \ce{Mn} lattice hosts magnetism with large $S=5/2$ \ce{Mn} moments.  The calculated magnetic moment of 4.5 $\mu_{\rm B}$/\ce{Mn} agrees well with the experimental $\mu_{\rm eff}$, indicating the predicted \ce{Mn}-moment between 4.04 and 5 $\mu_{\rm B}$ at the atomic limit~\citep{PhysRevMaterials.3.064202,zeugner2019chemical,li2020antiferromagnetic}. There is no change in the calculated moment in the 2D limit of the bilayer. The intralayer ferromagnetic (FM) coupling is described by FM superexchange coupling through the Goodenough-Kanamori-Anderson mechanism as the \ce{Mn-Te-Mn} bond angle is close to 90 degrees. Further, the FM layers are coupled antiferromagnetically, with the spins pointing in the out-of-plane direction~\citep{PhysRevLett.122.107202,PhysRevLett.124.167204,PhysRevB.99.155125,PhysRevResearch.1.012011,PhysRevX.9.041038,PhysRevMaterials.3.064202,PhysRevLett.122.206401}. Calculations favor the $\rm A$-type AFM structure, with energy difference $\Delta E_{\rm A/F}^{\perp}$ of $-$1.44 meV/\ce{Mn} for the bulk and $-$0.7 meV/\ce{Mn} for the bilayer compared to the corresponding FM state. The reduction of $\Delta E_{\text{A/F}}^{\perp}$ in the bilayer limit is attributed to the reduced coordination number in the out-of-plane direction.

To elucidate the magnetic interactions, we extract exchange couplings by mapping first-principles total energies onto an anisotropic Heisenberg Hamiltonian. In bulk \ce{MnBi2Te4}, the in-plane magnetization is primarily governed by the intralayer nearest-neighbor FM interaction $J_1$, while interactions beyond the first neighbor are at least one order of magnitude smaller (Table~\ref{tab:tab1}). The out-of-plane interaction is antiferromagnetic, $J_{\perp} = -0.04$ meV. Positive magnetic anisotropy, $A_z = 0.184$ meV, indicates an out-of-plane easy axis of magnetization. Classical Heisenberg Monte Carlo simulations predict a N\'eel temperature of 30 K in bulk \ce{MnBi2Te4} (Table~\ref{tab:tab1} and Supplemental Material~\citep{supple}), which is in good agreement with experimental findings and earlier theoretical results~\citep{otrokov2019prediction,PhysRevLett.122.107202,PhysRevLett.124.167204,PhysRevB.99.155125,PhysRevResearch.1.012011,PhysRevX.9.041038,PhysRevX.11.011003}. As expected, the N\'eel temperature decreases to 25.5 K in the bilayer, primarily due to the loss of neighbouring interactions in the perpendicular direction, and a reduction in $A_z$, which results in an increase in thermal spin fluctuations. 

While the $\rm A$-AFM magnetic ground state breaks the TR symmetry $\altmathcal{T}$, the combined symmetry $\altmathcal{S}=\altmathcal{T}u_{1/2}$ is preserved. Here,  $u_{1/2}$ is the half-translation operator connecting the nearest spin-up and spin-down \ce{Mn} layers. Similar to TR invariant topological insulators, $\altmathcal{S}$ leads to a topologically nontrivial electronic structure in \ce{MnBi2Te4} with $Z_2=1$ on the Brillouin zone plane, where $\mathbf{k}\cdot\mathbf{u}_{1/2} =0$ (Supplemental Material~\citep{supple}).  Additionally, the symmetry $\altmathcal{S}'=\altmathcal{P}_{\circ}\altmathcal{T}$ is preserved, resulting in spin-degenerate bands. Here, $\altmathcal{P}_{\circ}$ represents the spatial inversion symmetry centred around the point $O$ within the vdW gap [Figure~\ref{fig:fig1}(a)]. The valence bands are primarily composed of  \ce{Te} $p$-orbitals, while the conduction bands are dominated by \ce{Bi} $p$-orbitals [Figure~\ref{fig:fig1}(c)]. When the spin-orbit coupling is introduced, a band-inverted topological gap emerges at the $Z$-point (200 meV, Supplemental Material~\citep{supple}) and $\Gamma$-point [215 meV, Figure~\ref{fig:fig1}(a)]. The results align well with previous theoretical predictions and angle-resolved photoemission spectroscopy (ARPES) data~\citep{li2019intrinsic,PhysRevLett.122.107202,PhysRevX.9.041038,otrokov2019prediction}. Wilson-loop calculations confirm that \ce{MnBi2Te4} is an AFM topological insulator (Supplemental Material~\citep{supple}). The surface breaks the combined symmetry $\altmathcal{T}u_{1/2}$. Along with the out-of-plane magnetization, this symmetry breaking leads to the opening of a Dirac gap at the surface, significantly affecting the electronic properties.  The calculated magnetically induced surface gap of 26 meV is consistent with experimental measurements, which range between 50 and 100 meV as determined by ARPES and point-contact tunneling spectroscopy~\citep{otrokov2019prediction,li2019intrinsic,PhysRevLett.122.206401,zeugner2019chemical,acsnano.1c03936}.
In contrast, ferromagnetic bulk breaks both $\altmathcal{T}$ and $\altmathcal{P}_{\circ}\altmathcal{T}$ symmetries, resulting in a topologically nontrivial Weyl semimetal state (Supplemental Material~\citep{supple}), while the magnetic thin films exhibit the QAH effect.   

Topological states in vdW \ce{MnBi2Te4} can be tuned by quantum size effects. In films with even SLs, the combined $\altmathcal{S}'$ symmetry is preserved, ensuring double degeneracy of the bands. Conversely, both  $\altmathcal{T}$ and $\altmathcal{S}'$ symmetries are broken in odd-SL films, leading to spin-split bands. Even- and odd-layered films both exhibit magnetically induced surface gaps due to broken $\altmathcal{T}$ symmetry [Figure~\ref{fig:fig1}(e) and ~\ref{fig:fig1}(g)]. When the Fermi level is tuned to this gap, the surfaces demonstrate half-quantized Hall conductances of $\sigma_{xy}=\pm\ e^2/2h$, depending on the up and down orientation of the \ce{Mn}-spins at the surface. In the odd-layered films, the $\sigma_{xy}$ from two surfaces add up, leading to a  QAH insulating state with $\altmathcal{C} = 1$. The QAH effect has been experimentally observed in odd-SL thin flakes, achieving a Hall resistance of $R_{xy}=0.97\ h/e^2$ at zero magnetic field and 1.4 K in 5-SL flakes~\citep{deng2020quantum}. When an external magnetic field is applied to align the spins in all the SLs in parallel, both the quantization fidelity and critical temperature exhibit significant improvement. Accordingly, in even-layered AFM \ce{MnBi2Te4}, the Hall conductances from the two surfaces cancel each other, resulting in a zero-plateau QAH insulator. 

The characteristic band structure of the AFM bilayer \ce{MnBi2Te4} closely resembles that of the bulk, with a reduced topological gap of 82 meV [Figure~\ref{fig:fig1}(d)], consistent with previous calculations~\citep{PhysRevLett.122.107202}.  For the AFM 2SL phase, a gapped edge spectrum with a gap of $7.5 \text{ meV}$ is observed, with no evidence of a chiral edge mode [Figure~\ref{fig:fig1}(e)].  The combined $\altmathcal{PT}$ symmetry enforces local Berry curvature cancellation at every momentum point, $\Omega_{\rm top}(\mathbf{k}) = -\Omega_{\rm bottom}(\mathbf{k})$, ensuring a vanishing AHC, regardless of the $E_{\rm F}$ position [Figure~\ref{fig:fig1}(f)]. Shifting $E_{\rm F}$, however, fundamentally alters the nature of this zero-Hall response. When $E_{\rm F}$ lies within the gap, the zero-Hall plateau is explicitly topological. Here, the calculated layer-projected AHC is half-quantized to $\sigma_{xy}^{\rm t/b} = \pm(1/2)e^2/h$ (Supplemental Material~\citep{supple}) a hallmark of the axion insulating state ($\altmathcal{C}=0$) that is consistent with experimental observations~\citep{s41586-021-03679-w}.  Conversely, when the $E_{\rm F}$ shifts into the valence or conduction bands, the total AHC still converges to zero due to the persistent $\altmathcal{PT}$ symmetry cancellation. In this metallic phase, no quantized topological invariant is defined, and the layer-projected response deviates from $\pm(1/2)e^2/h$ due to the lack of a global energy gap.


When the interlayer coupling transitions to FM, the topological gap decreases to 51 meV, and the chiral edge state emerges within the bulk gap [Figure~\ref{fig:fig1}(g)]. The net Berry curvature now integrates to a nonzero Chern number, and the Hall conductivity becomes quantized, establishing a QAH insulating phase with $\altmathcal{C}$=1 [Figure~\ref{fig:fig1}(h)]. 

\begin{figure}[!t] 
\includegraphics[scale=0.11]{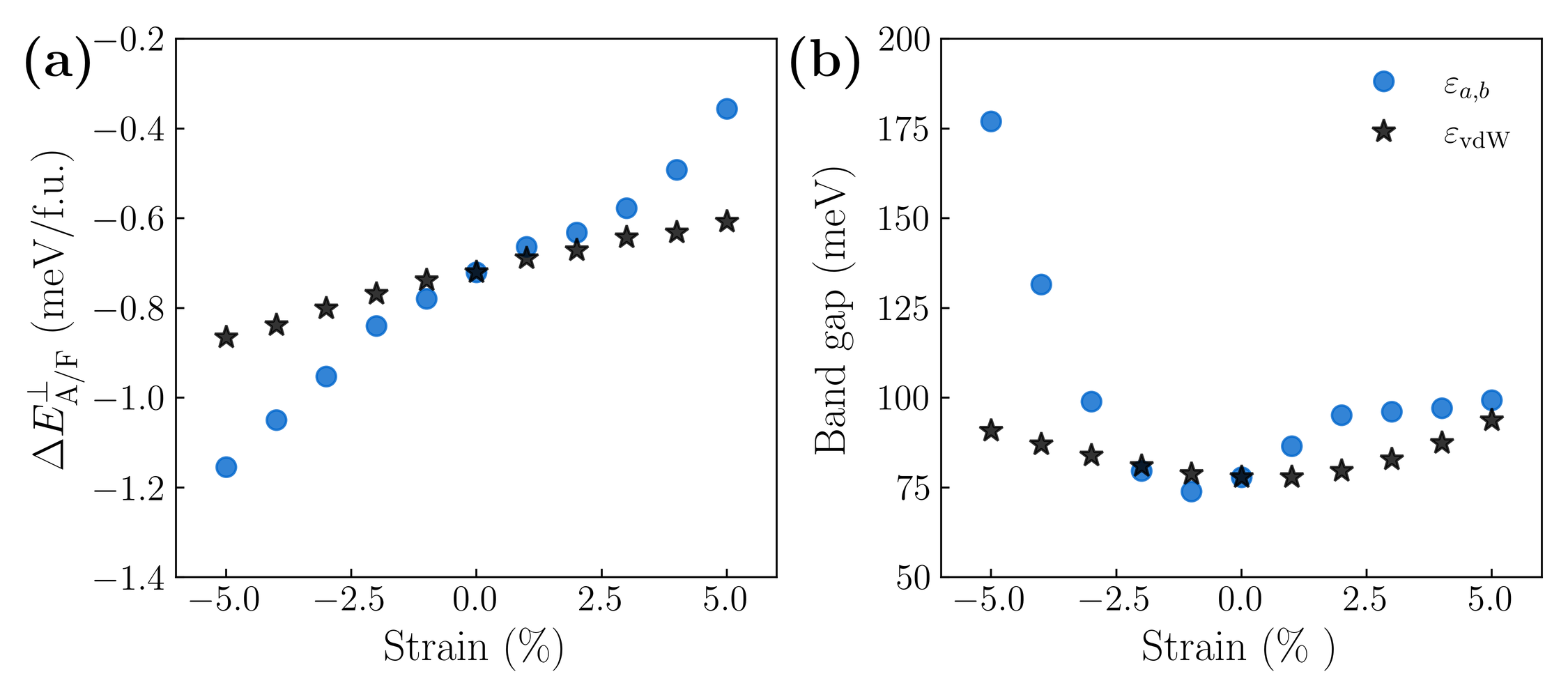}
\caption{Lattice strain does not induce magnetic or electronic phase transitions in bilayer \ce{MnBi2Te4}. (a) The interlayer exchange coupling remains AFM ($\Delta E_{\rm A/F}^{\perp} < 0$) under both in-plane biaxial strain $\varepsilon_{a,b}$ and variations in the vdW gap $\varepsilon_{\rm vdW}$. (b) The corresponding electronic structure remains gapped throughout. Consequently, the undoped bilayer retains its Axion insulating state, indicating that lattice strain alone is an insufficient perturbation to drive a transition to the conventional anomalous Hall regime. 
} 
\label{fig:strain} 
\end{figure}

\subsection{\label{resultsB}Manipulating Fermi-level and interlayer magnetic coupling}
A central question in ultrathin \ce{MnBi2Te4} is whether the interlayer AFM coupling can be tuned to FM order through external or chemical perturbations, and how the electronic structure evolves across this transition. In this context, external strain serves as a particularly clean control parameter, as it modifies orbital overlaps without introducing chemical disorder.  For instance, pressure-induced magnetic transitions in bilayer \ce{CrI3} have been achieved through stacking order modification~\citep{s41563-019-0506-1}, and strain-dependent magnetic order has been observed in metallic \ce{CrTe2} monolayer~\citep{s41563-026-02537-2}. Consequently, we investigate the 2SL by applying in-plane biaxial strain $\varepsilon_{a,b}$ and by modulating the vdW gap between the septuple layers, $\varepsilon_{\rm vdW}$ (Figure~\ref{fig:strain}). 

\begin{figure*}[!t] 
\includegraphics[scale=0.12]{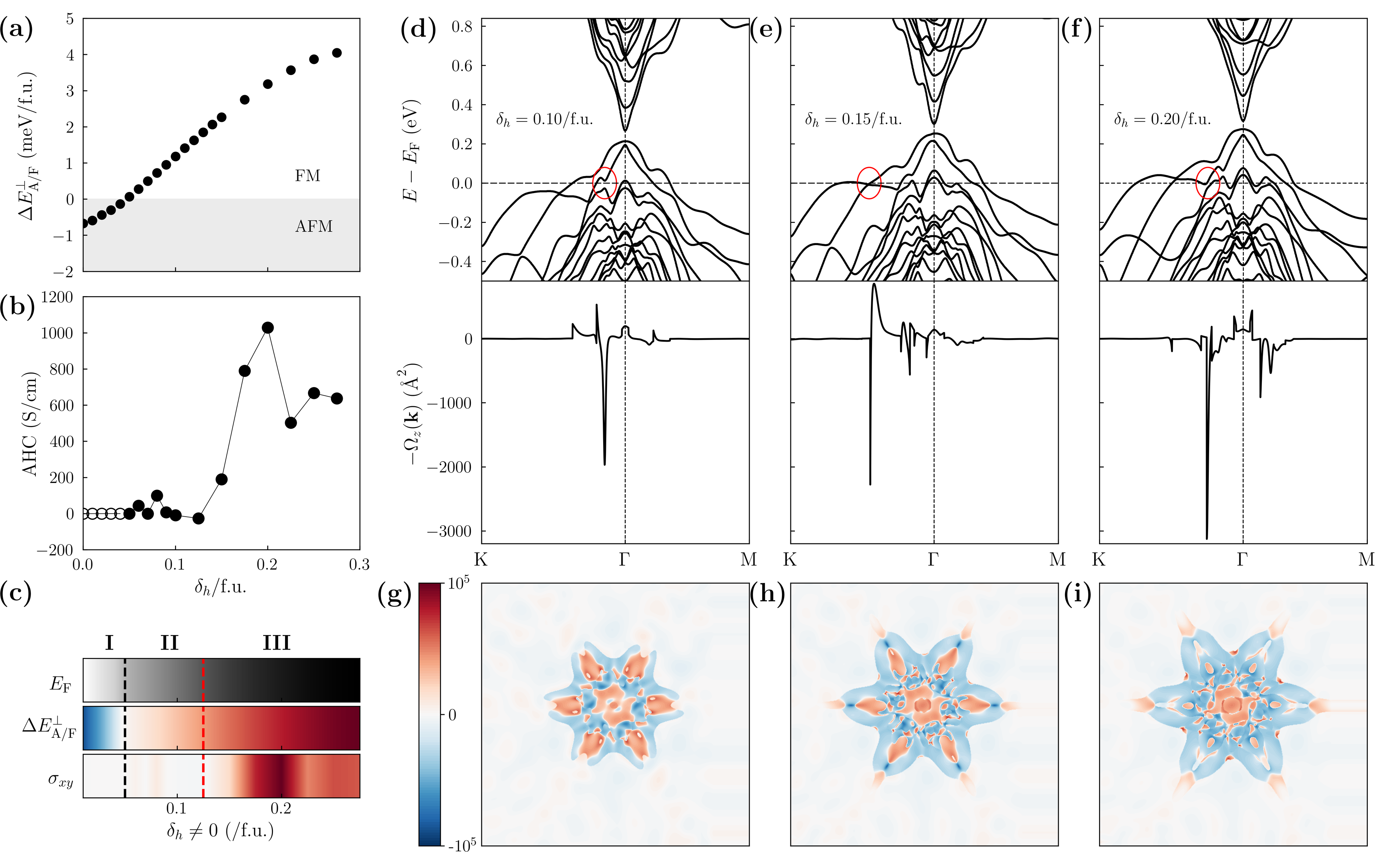}
\caption{
Magnetic and electronic evolution of bilayer \ce{MnBi2Te4} under hole doping. 
(a) The energy difference between the interlayer AFM and FM couplings, $\Delta E_{\rm A/F}^{\perp}$, in bilayer \ce{MnBi2Te4} indicates a transition from AFM to FM state as the doping concentration increases. 
(b) The calculated anomalous Hall conductivity peaks at a doping level of approximately 0.2 holes/f.u. doping. Open circles denote configurations where the interlayer coupling remains AFM.
(c) Schematic evolution of the electronic state (position of the Fermi level in the valence band from its maximum), magnetic state ($\Delta E_{\rm A/F}^{\perp}$) and transport properties ($\sigma_{xy}$), identifying three distinct regimes for $\delta_h \neq 0$. (I) Metallic AFM with $\sigma_{xy} =0$, (II) metallic FM with small $\sigma_{xy}$, and (III) metallic FM with giant $\sigma_{xy}$.
(d) - (f) The electronic structure of the FM 2SL under varying levels of hole doping reveals spin-orbit coupling-induced avoided band crossings near the Fermi level. These avoided crossings generate substantial Berry curvature $\Omega_z(\mathbf{k})$ at specific momentum points, contributing to the high anomalous Hall conductivity observed in the system. 
(g) - (i)  Momentum-resolved Berry curvature distribution, $-\Omega(\mathbf{k}, E_{\rm F})$, integrated over all occupied bands up to $E_{\rm F}$ corresponding to the doping levels in (d) - (f). The expansion of the Fermi surface with increasing doping is clearly visible across the 2D Brillouin zone. 
} 
\label{fig:fig2} 
\end{figure*}

Tensile $\varepsilon_{a,b} > 0$ strain quantitatively modifies the energy difference $\Delta E_{\text{A/F}}^{\perp}$ and the resulting $J_\perp$, yet it fails to induce a transition to interlayer FM order. Similarly, while increasing the vdW gap suppresses $\Delta E_{\rm A/F}^{\perp}$, the interlayer exchange $J_\perp$ remains AFM even under substantial strain. Conversely, compressive $\varepsilon_{a,b}<0$ strain further stabilizes the interlayer AFM coupling. Reducing the vdW gap, $\varepsilon_{\rm vdW} > 0$,  similarly strengthens the AFM  $J_\perp$,  which we attribute to the enhanced long-range superexchange across the vdW gap as orbital hopping improves with reduced distance [Figure~\ref{fig:strain}(a)]. These results are consistent with experimental observations of enhanced interlayer exchange coupling under hydrostatic pressure in thin films of \ce{MnBi2Te4}~\citep{acs.nanolett.5c05229} and \ce{MnBi2Te4(Bi2Te3)_n} family of materials~\cite{acs.nanolett.1c01874}.  Across all investigated strain regimes, the bilayer remains insulating, exhibiting a non-monotonic strain dependence of the electronic gap [Figure~\ref{fig:strain}(b)]. The robustness of both AFM $J_\perp$ and insulating gap suggests that mechanical strain is an ineffective perturbation for tuning $J_\perp$ order or shifting $E_{\rm F}$  into the bulk bands. Although we did not explicitly calculate $\sigma_{xy}$, the persistence of the AFM insulating state suggests that the native axion insulating phase, characterized by a zero-Hall plateau, is retained throughout.

In even-layer \ce{MnBi2Te4}, a large external magnetic field exceeding 9 \si{\tesla} can drive a transition from AFM to FM interlayer coupling, resulting in a quantum phase transition from the axion insulator to QAH state. This has been demonstrated in 6SL thin films~\citep{liu2020robust}. However, such high field requirements limit practical utility, and electrostatic gating offers a more scalable and device-compatible alternative. Indeed, gating has been established as an effective means of tuning interlayer exchange coupling and enhancing magnetic transition temperatures in ultrathin magnets~\cite{huang2018electrical, deng2018gate,PhysRevB.103.214411,PhysRevMaterials.6.084407,C9NR10171C}. Beyond magnetic tuning, gating shifts $E_{\rm F}$ into the bulk bands, where Berry curvature contributions from the partially filled bands are bounded by the Fermi surface rather than the full Brillouin zone.  This suppresses the quantization of the chiral Hall conductivity, and the system crosses over from the topological to the conventional anomalous Hall regime (Figure~\ref{fig:fig2}).    

Electrostatic gating can introduce either electrons or holes into the system, with the carrier density controlled by the applied voltage. We first investigate the scenario where the 2SL-\ce{MnBi2Te4} is hole-doped. Since the bands below the chemical potential are primarily composed of \ce{Te} $p$-orbitals, hole doping depletes electrons from these states,  shifting $E_{\rm F}$ into the valence manifold. Consequently, $E_{\rm F}$ is continuously tuned by the hole density $\delta_h$~\citep{supple}, the system becomes metallic, and the partially filled bands preclude the possibility of realizing a QAH state.  

Continuous depletion of \ce{Te}-$p$ electrons across the vdW gap gradually weakens the long-range interlayer AFM interaction, eventually driving a transition to FM order at a critical hole concentration of approximately 0.055 holes/f.u., which corresponds to \SI{3.34E13}{holes\per\square{\centi\meter}}. This transition is reflected in the evolution of the computed energy difference, $\Delta E_{\rm A/F}^{\perp}$, between the interlayer AFM and FM configurations [Figure~\ref{fig:fig2}(a)]. The interlayer AFM order in this system originates from a long-range Anderson superexchange mechanism~\citep{PhysRev.115.2}. Further analysis indicates that the Mn-$d_{z^2}$ orbitals interact across the vdW gap, mediated by the overlapping \ce{Te}-$p_z$-\ce{Bi}-$p_z$-\ce{Te}-$p_z$ orbitals~\citep{PhysRevLett.122.107202,PhysRevB.102.081107,supple}. Since the strength of this coupling is governed by the degree of delocalization of these $p_z$-orbitals, the depletion of \ce{Te}-$p$ electrons with increasing $\delta_h$  disrupts the AFM superexchange pathways (Supplemental Material~\citep{supple}). Consequently, interlayer FM coupling becomes increasingly favorable with hole doping, reaching $J_{\perp}=$ 0.153 meV at 0.2 hole/f.u. (Table~\ref{tab:tab1}). Similar to the bulk and undoped 2SL, the intralayer magnetism in the hole-doped bilayer remains predominantly governed by the nearest-neighbor FM coupling $J_1$, with higher-order terms remaining small ($J_2/J_1, J_3/J_1 < 0.1$). However, $J_1$ increases significantly to 0.453 meV, driven by induced metallicity. The finite density of states at the $E_{\rm F}$ opens additional interaction channels via itinerant conduction electrons, thereby enhancing the effective intralayer coupling. Combined with the strong FM $J_\perp$, these enhanced interactions yield a substantially higher Curie temperature $T_{\rm C}$ of 60 K, as determined from the Heisenberg Monte Carlo simulations (Table~\ref{tab:tab1} and Supplemental Material~\citep{supple}).

Induced metallicity drives the system away from the topological quantum phases and into a conventional Hall regime.  This is consistent with the conductivity tensor evolution observed experimentally when $E_{\rm F}$ is tuned away from the QAH plateau via gating in ferromagnetic doped topological insulators~\citep{nphys3053}. Below a critical hole density, the computed AHC is nearly zero, and as $\delta_h$ increases, intrinsic $\sigma_{xy}$ rises monotonically, reaching a peak value of 1127 S/cm  near 0.2/f.u.\ density [Figure~\ref{fig:fig2}(b)], a remarkably large AHC for a 2D magnet. This value is comparable to that of the ferromagnetic kagome semimetal  \ce{Co3Sn2S2}~\citep{s41567-018-0234-5,s41467-018-06088-2}, and substantially exceeds those reported for the topological nodal-line semimetals \ce{Fe3GeTe2}~\citep{s41563-018-0132-3} and \ce{MnAlGe}~\citep{adma.202006301}, the ferromagnetic kagome metal \ce{Fe3Sn}~\citep{Kida_2011,10.1038/nature25987}, and the itinerant ferromagnet bcc-\ce{Fe}~\citep{PhysRevB.105.115135}. To place this in a broader context, we compare the AHC across flakes of varying thickness. The monolayer 1SL is a trivial insulator, yet hole-doping exhibits a dome-shaped variation in  $\sigma_{xy}$, yielding $\sigma_{xy} \sim 400$ S cm$^{-1}$ at 0.2/f.u.\ hole density (Supplemental Material~\citep{supple}). This is consistent with the estimate $e^2/hd_z \sim 277$  S/cm, where $d_z \sim 1.4 $ nm is the thickness of the uncompensated magnetic layer in odd-layer films~\citep{deng2020quantum,s41467-022-29259-8}. Furthermore, in qualitative agreement with the present results, experimental measurements on 7SL films show a gate-tunable  $\sigma_{xy}$ reaching $\sim$ 550 S/cm~\citep{s41467-022-29259-8}, lending support to the present calculations.

\begin{figure*}[!t]
\includegraphics[scale=0.2]{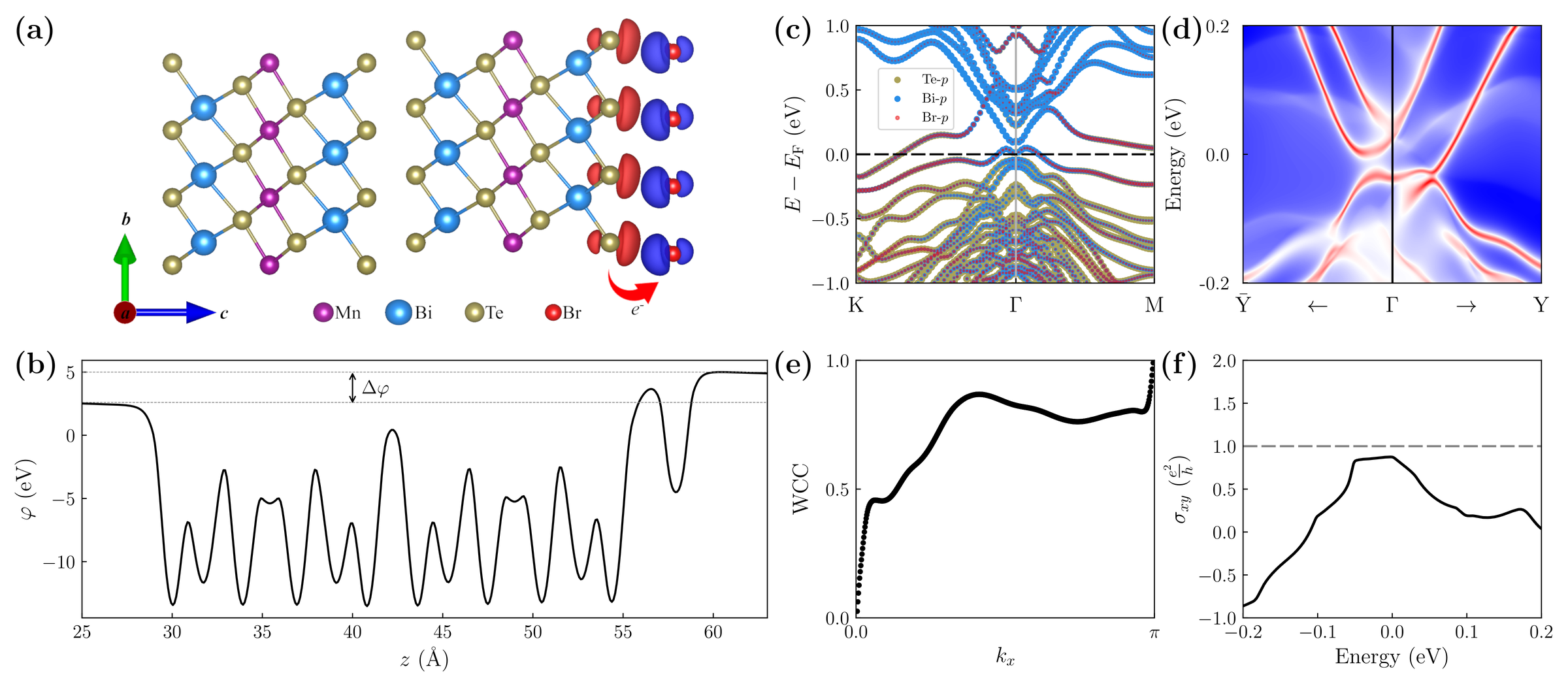}
\caption{
Surface functionalization and topological transport in Br-doped bilayer \ce{MnBi2Te4}. (a) Schematic illustration of the 2SL structure with electronegative \ce{Br} atoms absorbed on the top surface. Bonding charge density analysis, $\Delta\rho(\mathbf{r})$, reveals electron depletion (red) and accumulation (blue) at the \ce{Te} and \ce{Br} sites, respectively, confirming that surface functionalization effectively hole-dopes the system. 
(b) Calculated electrostatic potential $\varphi$, perpendicular to the two-dimensional plane. A significant potential difference, $\Delta\varphi = $  2.5 eV, indicates a strong internal electric field that breaks the inversion symmetry. 
(c) Calculated electronic structure exhibiting metallic behavior with significant hybridization between the \ce{Te}-$p$ and \ce{Br}-$p$ orbitals. 
(d) The boundary spectral function displaying a topological edge mode.
(e) The calculated Wannier charge center (WCC) confirms its topological nature, indicating a Chern number of $\altmathcal{C} = 1$. 
(f) The corresponding anomalous Hall conductivity indicates a breakdown in exact quantization due to interfering bulk bands with the topological edge state. 
}
\label{fig:fig3}
\end{figure*}

For $\delta_h \neq 0$, the electronic structure is inherently metallic as the Fermi level is progressively pushed into the valence band with increasing hole density (Supplemental Material~\citep{supple}). Within this metallic state, three distinct transport regimes emerge that warrant microscopic attention [Figure~\ref{fig:fig2}(c)]. In the first regime, characterized by low hole doping, the system behaves as an AFM metal with $\sigma_{xy}=0$.  Here, the Berry curvature contributions from the AFM coupled layers exactly cancel by symmetry, yielding a vanishing net AHC. As doping increases into the second regime, the system transitions to an FM-metal phase, though $\sigma_{xy}$ remains notably small. Finally, in the third regime, this same FM-metallic phase hosts a colossal $\sigma_{xy}$, reaching a peak value of 1127 S/cm.

To elucidate the microscopic origin of this behavior, we computed the band structures, the Berry curvature along high-symmetry $k$-paths, and the Berry curvature distribution over the 2D BZ for all occupied bands up to $E_{\rm F}$ at representative doping levels [Figure~\ref{fig:fig2}(d)-(i)]. Spin-orbit coupling induces avoided crossings between bands at specific momentum points, generating a substantial Berry curvature when the Fermi level is located near such an anticrossing point. However, the presence of these hot spots alone is insufficient to produce a large AHC, as the net $\sigma_{xy}$ is determined by the integral of the Berry curvature over the Fermi sea, and the local hot spots of opposite sign cancel.

In the FM-metal phase below 0.12 holes/f.u.\, the positive and negative $\Omega(\mathbf{k},E_{\rm F})$ contributions nearly compensate for one another, resulting in $\sigma_{xy} \sim 0$. At higher doping levels, this compensation breaks down, yielding a net Berry curvature. Simultaneously, the Fermi surface expands ($\propto k_{\rm F}^2$), as $E_{\rm F}$ is pushed deeper into the valence band. This results in an increased Fermi momentum $k_{\rm}$ as $\delta_h$ rises. The resulting net Berry curvature, accumulated over the enlarged Fermi sea, drives the intrinsic AHC to its peak value and remains high for a broad doping range. This microscopic picture provides a unified explanation for the non-monotonic $\sigma_{xy}$ observed across three regimes. Furthermore, it underscores that gate-tuneable AHC in 2SL \ce{MnBi2Te4} is a direct consequence of the interplay between interlayer FM coupling, Fermi surface topology, and Berry curvature redistribution under hole doping. 

It is worth noting that in many magnetic systems, large $\sigma_{xy}$ is predominantly extrinsic in origin, driven by impurity-induced skew-scattering and side-jump mechanisms~\citep{RevModPhys.82.1539, sciadv.abb6003, acsnano.1c00488}. Such extrinsic contributions are strongly temperature dependent and are typically observable only well below the magnetic ordering temperature. In contrast, the giant AHC in hole-doped 2SL is intrinsic, originating from uncompensated Berry curvature hot spots, and is expected to persist up to the magnetic ordering temperature. The significantly enhanced $T_{\rm C}$ (Table~\ref{tab:tab1}) in hole-doped flakes further extends the temperature window over which this large AHC may be observed. 

We now investigate the electron-doping scenario, where electrostatic gating shifts $E_{\rm F}$ into the conduction band, driving an insulator-to-metal transition.  In contrast to hole doping, the interlayer coupling remains AFM under electron doping (Supplemental Material~\citep{supple}). This marked electron-hole asymmetry in the magnetic response directly reflects the orbital-selective nature of the superexchange mechanism. In contrast to the primary hole distribution over the \ce{Te}-$p$ orbitals, the added electrons are distributed evenly across the $p_x$, $p_y$, and $p_z$ orbitals of both \ce{Bi} and \ce{Te}, only mildly perturbing the interlayer exchange without altering its AFM character.  Since the AFM interlayer coupling enforces compensation of Berry curvature contributions from the two layers, $\sigma_{xy}$ remains negligible across the electron-doped regime. This marked electron-hole asymmetry in the Hall response highlights the orbital selective nature of magnetic exchange in \ce{MnBi2Te4}.

Chemical doping provides an alternative route for carrier injection, and we explore whether it can replicate the magnetic phase transition achieved under electrostatic gating. Motivated by the efficacy of hole doping, we employ electronegative \ce{Br} adatoms to withdraw electrons from the bilayer. By systematically placing \ce{Br} atoms at various high-symmetry sites, we identify the lowest-energy configuration to be the center of the $ab$-plane of the unit cell. Bonding charge density analysis, $\Delta \rho(\mathbf{r})$, confirms that the \ce{MnBi2Te4} surface loses electrons upon \ce{Br} absorption, establishing that this surface functionalization effectively hole-dopes the system [Figure~\ref{fig:fig3}(a)]. Furthermore, the one-sided functionalization introduces an internal electric field that breaks the inversion symmetry of the bilayer, as evidenced by a large electrostatic potential difference of $\Delta\varphi$ = 2.5 eV between the two surfaces [Figure~\ref{fig:fig3}(b)]. 

At full \ce{Br} coverage, the interlayer coupling transitions to FM order, with an energy difference of $\Delta E_{\rm A/F}^{\perp}$ = \SI{1.64}{\milli\electronvolt}/f.u., corresponding to an FM exchange of $J_\perp$ of 0.088 \si{\milli\electronvolt} (Table~\ref{tab:tab1}). A microscopic analysis of exchange interactions reveals a picture qualitatively distinct from the neutral and gated cases (Table~\ref{tab:tab1}).  While the in-plane magnetism in bulk and bilayer is primarily governed by the FM $J_1$, the longer-range $J_2$ and $J_3$ exchanges become significant upon \ce{Br} functionalization (Table~\ref{tab:tab1}). Consequently, a $J_1$-$J_2$-$J_3$ Heisenberg model is required to accurately describe the magnetic behavior, and the predicted Curie temperature increases significantly to 54 K. 

The metallic character of the \ce{Br}-functionalized bilayer is qualitatively different from that of the hole-doped case, arising from strongly hybridized \ce{Br}-$p$ orbitals with the \ce{Te}-$p$ states [Figure~\ref{fig:fig3}(c)], localized at the functionalized surface. The two bands crossing the Fermi level exhibit hole-pockets at the $\Gamma$-point and an electron pocket at the $K$-point. Importantly, band inversion persists at the $\Gamma$-point on both sides of $E_{\rm F}$, buried beneath these functionalization-induced metallic bands. This suggests that topological character survives despite the induced metallicity. Indeed, the boundary spectral function from the Green's function method reveals a chiral edge mode [Figure~\ref{fig:fig3}(d)], and the Wannier charge center analysis of the occupied bands yields $\altmathcal{C}=1$  [Figure~\ref{fig:fig3}(e)], establishing the topological character of the state. 

The system lacks a global gap, leading to spectral overlap between the chiral edge mode and the functionalization-driven metallic bands of the bilayer. Consequently, the occupied-state Berry curvature no longer integrates cleanly to an integer Chern number. Instead, the partially filled metallic bands contribute an additional, non-quantized Berry-curvature term, yielding a reduced $\sigma_{xy} \sim 0.86\ e^2/h$ [Figure~\ref{fig:fig3}(f)] that exhibits a weak residual $E_{\rm F}$ dependence rather than a flat plateau. This residual slope is itself characteristic of a Fermi sea contribution, where the shifting $E_{\rm F}$ alters the underlying Fermi surface topology. In a complementary transport picture, the same metallic bands furnish inter-edge backscattering channels that couple the counter-propagating chiral modes on opposite boundaries, introducing dissipation and degrading exact quantization. The resulting incomplete quantization is consistent with prior studies showing that the coexistence of chiral edge modes with conducting bulk states, whether arising from disorder or thermally activated carriers, disrupts exact QAH quantization by providing backscattering pathways~\citep{pnas.1424322112, s41467-021-25912-w,PhysRevB.105.165411}. In the \ce{Br}-functionalized bilayer, the functionalization-derived metallic bands play the analogous role.

\section{\label{summary}Summary}
 In summary, we systematically investigate the evolution from the topological to the conventional anomalous Hall regime in ultrathin \ce{MnBi2Te4} via electrostatic gating and controlled surface functionalization. We demonstrate that hole doping tunes the interlayer antiferromagnetic coupling to ferromagnetic order, while the concurrent insulator-to-metal transition triggers robust anomalous Hall transport, with conductivity reaching 1127 S/cm, among the highest reported in quantum materials. Microscopic analysis reveals that the complex evolution of the anomalous Hall conductivity with hole density arises from a delicate interplay among interlayer exchange coupling, Fermi surface topology, and the redistribution of Berry curvature. In contrast, surface-specific chemical \ce{Br} functionalization introduces an asymmetric charge distribution that generates an internal vertical electric field, thereby breaking the structural inversion symmetry. In this regime, the chiral edge mode coexists spectrally with functionalization-induced metallic bulk states, and the absence of a global gap displaces the intrinsic $\sigma_{xy}$ to $\sim 0.86\ e^2/h$. While electrostatic gating and chemical doping yield distinct transport signatures, both routes consistently enhance the effective intralayer ferromagnetic exchange interactions. This results in a substantial elevation of the magnetic ordering temperature in both cases, as determined by Heisenberg Monte Carlo simulations. Collectively, these results establish a dual-control framework for simultaneously engineering the topological, magnetic, and transport properties in ultrathin \ce{MnBi2Te4}, and motivate further experimental investigations.

\section{Acknowledgements}
B. G. acknowledges a fellowship from the University Grants Commission. We acknowledge the support and resources provided by the PARAM Brahma Facility at IISER, Pune, under the National Supercomputing Mission of the Government of India.  M. K. acknowledges funding from the National Mission on Interdisciplinary Cyber-Physical Systems of the Department of Science and Technology, Government of India, through the I-HUB Quantum Technology Foundation, Pune, India.


%

\clearpage
\onecolumngrid

\newcounter{pdfpagecount}
\setcounter{pdfpagecount}{1}

\makeatletter

\whiledo{\value{pdfpagecount} < 8}{
    \thispagestyle{empty} 
    \noindent
    \centerline{
        \includegraphics[page=\value{pdfpagecount}, width=0.95\paperwidth, height=0.95\paperheight, keepaspectratio]{Supplemental_Material.pdf}
    }
    \clearpage
    
    \ifnum\value{pdfpagecount}=7
        \let\ps@plain\ps@empty
        \let\ps@headings\ps@empty
        \thispagestyle{empty}
    \fi
    
    \stepcounter{pdfpagecount}
}

\makeatother

\pagestyle{empty}

\end{document}